\documentclass[10pt]{aipproc}

\layoutstyle{6x9}

\usepackage{amssymb}
\usepackage{latexsym}

\begin{document}

\newcommand{\U}[1]{\,\mathrm{#1}}
\newcommand{\gcmcm}{\U{g} \U{cm}^{-2}}
\newcommand{\epsfile}[1]{#1.eps}

%
%

\title{Detection of inclined and horizontal showers in the Pierre Auger Observatory}
\author{V. Van Elewyck \\ {\small for the Pierre Auger Collaboration}}{address={Observatorio 
Pierre Auger, Av. San Martin Norte 304, (5613) Malarg\"{u}e, 
Argentina},email={vero@nucleares.unam.mx}}

\begin{abstract}
The Pierre Auger Observatory can detect with high efficiency the air showers induced by ultra-high 
energy cosmic rays incident at large zenith angles $\theta > 60^\circ$. We describe here the 
specific characteristics of inclined and horizontal showers, as well as the characteristics of 
their signal in the surface detector. We point out their relevance both to extend the potential of 
the detector, and in the context of the detection of high-energy cosmic neutrinos.   
\end{abstract}
\maketitle

\section{Introduction}

The Pierre Auger Observatory (PAO) \cite{pao97} is a hybrid detector designed for the study of the 
energy spectrum, direction and composition of cosmic rays above $10^{19}$ eV. The surface detector 
uses water Cherenkov tanks to detect the footprint of the extensive air showers induced by the 
interaction of the cosmic rays in the atmosphere. This technique makes the PAO suited for the 
detection of inclined showers, with zenith angle $\theta > 60^\circ$.

The region of zenith angle $60^\circ \leq \theta \leq 90^\circ$ contributes half the total solid 
angle and 25\% of the total geometrical acceptance of the detector. It thus significantly extends 
its field of view and provides an opportunity to complement the studies of extensive air showers 
realised in the range $0^\circ \leq \theta \leq 60^\circ$. 

Moreover, the possibility of detecting deeply penetrating, very inclined showers could also enable 
the PAO to operate as a high energy neutrino detector 
\cite{Capelle:1998zz,bertou2002,Aramo:2004pr}. Neutrino induced showers would have to be 
identified and separated from a background of inclined showers produced by protons or nuclei. The 
study of this background has lead to the development of new techniques to analyze this type of 
showers \cite{ave00}.

\section{Characteristics of inclined air showers}
\begin{figure}[bt]
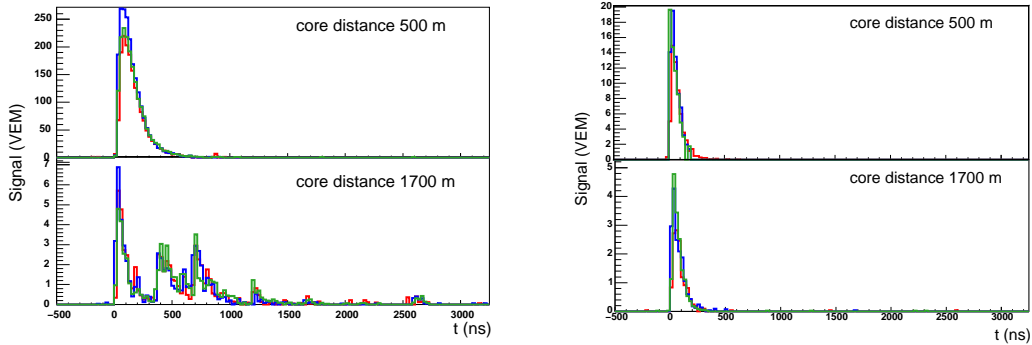

  \begin{minipage}[c]{0.49\linewidth}
  \centering
  \includegraphics[width=0.90\textwidth]{\epsfile{FADC-791081-merged}}
  \end{minipage}
  \hfill
  \begin{minipage}[c]{0.49\linewidth}
  \centering
  \includegraphics[width=0.85\textwidth]{\epsfile{FADC-759790-merged}}
  \end{minipage}
  \caption{FADC traces of a young (left) and old (right)
    shower. We see on the left how the signal gets smaller and more
    extended in time as the core distance changes from \(500\U{m}\) (top left)
    to \(1700\U{m}\) (bottom left). Old showers have short traces
    at all core distances (right).}
  \label{fig:FADC}
\end{figure}

Normal cosmic rays initiate an air shower in the first few $100 \gcmcm$ of the atmosphere. The 
composition and time structure of the shower at ground depend on the distance between the shower 
production point and the detector position, which increases with the zenith angle. 
Studies of the development of inclined showers from nucleonic primaries show that their 
electromagnetic component from $\pi^0$
production and decay dies out before reaching the ground 
 \cite{Zas:2005zz}. The only particles arriving at the detector are energetic muons (typically of 
$10-1000$ GeV), accompanied by an electromagnetic halo which is constantly regenerated by muon 
decay, bremsstrahlung and pair production, and contributes less than 15\% to the overall Cherenkov 
signal in the tank. Those muons arrive at the ground in a thin front with small curvature, 
resulting in short FADC pulses in the tanks (see fig.~\ref{fig:FADC}). 

The trajectories of the muons are extended enough to be affected by the geomagnetic field, which 
leads to a separation between positive and negative muons. In the plane perpendicular to the 
shower axis, the muon pattern gets extended and can even assume a lobed form for very inclined 
showers. This results in very elongated footprints in the ground plane, leading to events with 
high tank multiplicity. 

\begin{figure}[bt]
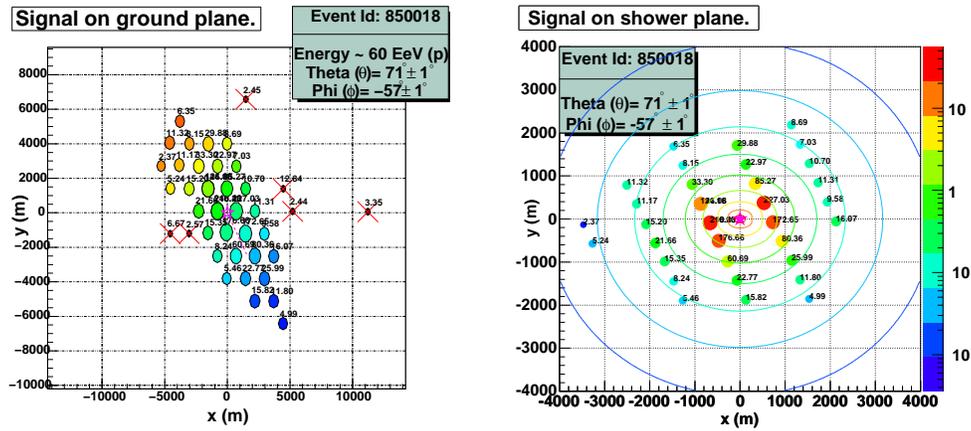

  \centering
  \includegraphics[width=0.38\textwidth, angle=270]{\epsfile{Event850018_Ground-fixed}}%
  \qquad
  \includegraphics[width=0.38\textwidth, angle=270]{\epsfile{Event850018_Shower}}
  \caption{A reconstructed, inclined event showing the surface detector data as seen on the ground
    (left) and in the plane orthogonal to the shower axis (right). One
    can see the distortions of the iso density lines due to 
    geomagnetic effects. The energy assignment of $\approx 60$ EeV assumes that the primary 
particle is a proton and that the hadronic interaction model is QGSJET. }
  \label{fig:event}
\end{figure}

This effect, together with the difference in evolution of the shower along the footprint on the 
ground, imply that for very inclined, old showers
the cylindrical symmetry around the shower axis is lost, as shown in fig.~\ref{fig:event}. A 
specific technique of reconstruction is required, based on the search for the best fit to the muon 
pattern on the ground, using an average shower characterized by the map of muon densities at 
ground and by the average response of a tank to incident muons. The knowledge of the best fitting 
muon map provides us with a characterisation of the shower which is, in good approximation, 
independent of the composition of the nuclear primary. On the contrary, the scale of the muon 
density, which can be used to infer the primary energy, changes with the composition; the details 
of this dependance is obtained using Monte Carlo simulations which are affected by our incomplete 
knowledge of interactions at high energies.

In this context, hybrid events reconstructed by both the surface and the fluorescence detector 
will play an important r\^{o}le in understanding inclined showers, and in particular their 
relation with the primary composition, since they allow independant measurements of the 
electromagnetic and muonic components of the shower \cite{ave03}.

\section{Neutrino induced showers}
 Due to their small cross-section, neutrinos can penetrate the atmosphere deeply and initiate 
showers at all possible depths, contrary to nuclear or electromagnetic primaries. In particular, 
showers originating less than $\approx 2000 \gcmcm$ away from the detector will reach it before 
their electromagnetic component attenuates completely. The corresponding signal is that of a young 
shower: short, peaked FADC traces close to the core position, and extended traces far from it, 
reflecting the large curvature of the shower front and the presence of an electromagnetic 
component. At large zenith angles, with these criteria it is in principle possible to separate a 
young, neutrino induced shower from an old shower induced by a nuclear primary (see 
fig.~\ref{fig:FADC}). The aperture of the PAO for neutrino induced showers should be comparable to 
that for contained events in conventional neutrino telescopes \cite{Capelle:1998zz}.

\section{Conclusions}
Through the study of inclined air showers, the PAO significantly extends its detection potential 
and becomes sensitive to another component of cosmic radiation, the neutrinos. The specific 
phenomenology of those showers also opens new ways for other analyses, like mass composition 
studies.

\end{document}